\def\be{\begin{equation}}
\def\ee{\end{equation}}
\def\ber{\begin{eqnarray}}
\def\eer{\end{eqnarray}}
\def\bers{\begin{eqnarray*}}
\def\eers{\end{eqnarray*}}

\documentclass[journal=jacsat,manuscript=article]{achemso}

\usepackage{dcolumn}   
\usepackage{amsmath}
\usepackage{multirow}
\usepackage{graphicx}    
\usepackage{subfigure}
\usepackage{comment}
\usepackage{color}
\usepackage{makecell}
\usepackage{ulem}
\usepackage[utf8]{inputenx}
\DeclareUnicodeCharacter{2212}{\-}
\usepackage{ifthen}
\newboolean{includefigs}
\setboolean{includefigs}{true}      
\newboolean{includetext}
\setboolean{includetext}{true}     
\newcommand{\condcomment}[2]{\ifthenelse{#1}{#2}{}}

\title{Double Half-Heusler Alloys X$_2$Ni$_2$InSb (X= Zr/Hf) with promising Thermoelectric Performance: Role of varying structural phases}
\author{ Bhawna Sahni and Aftab Alam}
\email{aftab@phy.iitb.ac.in}
\affiliation{Department of Physics, Indian Institute of Technology, Bombay, Powai, Mumbai 400 076, India}
\begin{document}
\begin{abstract}
Double half-heusler alloys are the new class of compounds which can be seen as transmuted version of two single half-heusler with higher flexibility of tuning their properties. Here, we report a detailed study of thermoelectric (TE) properties of two double half-heusler (HH) alloys X$_2$Ni$_2$InSb (X=Hf/Zr), using first-principles calculation. These alloys exhibit a rich phase diagram with the possibility of tetragonal, cubic and solid solution phase at different temperature range. As such, a comparative study of TE properties of all these phases is performed. The ordered phases show quite favorable electronic transport as compared to the disordered ones in both compounds. Lattice thermal conductivity of double HH alloys is lower than their ternary counterpart, making them most promising for TE application. Simulated band gap, obtained using hybrid functional, of ordered phases of Hf$_2$Ni$_2$InSb and Zr$_2$Ni$_2$InSb lie in the range 0.24-0.4 eV and 0.17-0.59 eV respectively, while for disordered phase, it is 0.05-0.06 eV. Hf$_2$Ni$_2$InSb shows a reasonably high ZT value of $\sim$ 2.19, while Zr$_2$Ni$_2$InSb yields 2.46 at high temperature for n-type conduction in tetragonal phase. The ZT value for p-type conduction is also quite promising ($\sim$ 1.35 and $\sim$ 2.19 for Hf- and Zr-based compounds). In both the compounds, electronic transport (Seebeck and electrical conductivity) plays the dominant role for the high ZT-value. Keeping in mind the promising TE performance, we propose immediate attention from experimentalists to synthesize and cross validate our findings for these new candidate materials.   
\end{abstract}

\section{Keywords}
	\textit{Double half-heusler alloys, Thermoelectric, Ab-initio calculation, Carrier relaxation time, Electronic and thermal transport}


\maketitle
\section{Introduction}
Thermoelectric technology which enables to convert waste heat into electricity, has proven to be extremely useful in providing solutions to renewable energy resources. Since most of the energy from primary sources is lost as waste heat, potential thermoelectric materials come as a rescue by harvesting this waste heat. Transport properties (Seebeck Coefficient (S), electrical conductivity ($\sigma$) and thermal conductivity ($\kappa$) which define the thermoelectric figure of merit (ZT), are closely interrelated, which makes it quite challenging to find novel materials with optimal ZT. Though there are several materials reported in the literature,\cite{key-1,key-2,key-3,key-4,key-5,key-6,key-7,key-8,key-9} the hunt for more efficient novel materials is still ongoing.

Half-Heusler (HH) alloys have emerged as promising thermoelectric materials due to a variety of interesting properties such as good thermal stability, easily tunable band gaps, good mechanical properties etc.\cite{APL79, APL86,acta.Mater.57,APL_92,JPCM01} They can be classified on the basis of the valence electron count (VEC). The 18 valence-electron HH alloys are very stable because of fully occupied bonding and empty anti-bonding states. The 17 and 19 valence-electron HH alloys on the other hand, are unstable because of partially occupied states. However, mixing of a 17 and a 19 VEC HH alloy can form an 18 VEC double half heusler alloy. These double half heusler alloys exhibit much lower values of lattice thermal conductivity ($\kappa_L$) as compared to their ternary counterpart because of smaller phonon group velocity and disorder scattering. Apart from $\kappa_L$, if the electronic transport properties of these double HH alloys can be made more superior than the corresponding ternary systems, they can be very promising for TE applications. This is one of the motivation of the present work.

Anand \textit{et al.}\cite{double_HH1} explored a large number of unexplored double half-Heusler alloys and predicted many of them stable. The double half-heusler compounds have a general formula unit X\textsubscript{2}YY$^{'}$Z\textsubscript{2} where Y and Y$^{'}$ are not isovalent, X is transition metal and Z is a main group element. For example, the nominal net valence (NV)$\neq$0 systems such as TiNiSb and TiFeSb are the two ternary components of the quaternary double half-Heusler compound, Ti$_2$FeNiSb$_2$. The members, namely TiFeSb, TiCoSb, and TiNiSb all have different net valence (NV = -1, 0, and 1). Anand \textit{et al.}\cite{double_HH1} showed that $\kappa_L$ of double half-Heusler Ti$_2$FeNiSb$_2$ is lower in comparison to that of its corresponding (NV = 0) ternary counterpart TiCoSb (with the same average atomic mass) by a factor of 3 at room temperature. There are some reports on the effect of doping in these double half heuslers as well. For instance, Liu \textit{et al.}\cite{Liu} showed that by alloying Ti by Hf and by tuning Fe/Ni ratio, a high figure of merit can be achieved for both p-type and n-type conduction in TiFe$_{0.5}$Ni$_{0.5}$Sb. Wang \textit{et al.} showed enhanced thermoelectric performance due to very low value of thermal conductivity and high power factor in p-type double half heusler Ti\textsubscript{2-y}Hf\textsubscript{y}FeNiSb\textsubscript{2-x}Sn\textsubscript{x}
compounds.\cite{dHH} Recently,  Hasan {et al.} \cite{dHH1} showed enhanced figure of merit in   Ti$_2$FeNiSb$_{1.8}$Sn$_{0.2}$ as compared to the pristine Ti$_2$FeNiSb$_2$.

Since the ternary half-heuslers ZrNiSn\cite{ZrNiSn,APL79,APL86} and HfNiSn\cite{APL86,HfNiSn} have been widely studied for reporting promising thermoelectric performance, we chose to study the corresponding (NV=0) double half heusler counterparts Zr$_2$Ni$_2$InSb and Hf$_2$Ni$_2$InSb. Anand \textit{et al.}\cite{double_HH1} theoretically  studied the Gruniesen parameter and thermal conductivity of these compounds in tetragonal (I-42d) structure as reported in the open quantum materials database (OQMD). Both these compounds, however, are experimentally reported to crystallize in cubic (F$\bar{4}$3m ($\#$216)) structure with few impurities, prepared under a specific processing condition.\cite{dHH_2}$^{,}$\cite{ZrNIInSb} 

In this letter, we use first-principles calculation to investigate the electronic, phonon and thermoelectric properties of double half heuslers Zr$_2$Ni$_2$InSb and Hf$_2$Ni$_2$InSb. In contrast to previous studies, we have simulated these properties in all of three relevant phases i.e., cubic, tetragonal as well as solid solution, of these compounds. Depending on the synthesis condition and temperature, there is a possibility to realize all these three phases, ordered in low temperature (T) while disordered in high T-range. Ordered structures could show high carrier mobility than solid solution.\cite{ordered} Thus, ordered structures with low lattice thermal conductivity ($\kappa_L$) can prove to be better thermoelectric material. For an accurate estimate of band gap, HSE06 functional is used, which yields a band gap of 0.17 eV (0.59 eV) for cubic (tetragonal) phase of Zr$_2$Ni$_2$InSb, while the same for Hf$_2$Ni$_2$InSb are 0.24 eV (0.4 eV). The corresponding solid solution phase shows a narrow band gap 0.05 eV and 0.06 eV for Zr and Hf based double half heusler alloys. As expected the simulated $\kappa_L$ for double half heusler alloys are smaller than the ternary counterpart (ZrNiSn and HfNiSn). However these values of $\kappa_L$ is not small enough to give promising TE properties. Interestingly, these double HH alloys are found to show extremely high electronic transport (S, $\sigma$ and power-factor), which actually makes them promising for TE application. Ordered phases are found to be almost equally competitive with respect to their TE performance with figure of merit (ZT) value as high as 2.46, at high T. Such comparative study of different structural phases of a single compound is extremely essential and useful to understand the nature of electrons and phonons excitation at different T-ranges.

\section{Computational Details}
We use Vienna Ab-initio Simulation Package (VASP)\cite{VASP1}$^,$\cite{VASP2}$^,$\cite{VASP3}, within DFT, with a projector augmented wave basis\cite{PAWbasis} and the generalized gradient approximation exchange-correlation functional of Perdew$-$Burke$-$Ernzerhof (PBE)\cite{PBE}.  HSE06 \cite{HSE06} calculations including spin-orbit coupling (SOC) were performed for the accurate estimation of band gaps. A plane-wave energy cutoff of 500 eV was used. The Brillouin zone sampling was done by using a $\Gamma$-centered $k$-mesh. For all the compounds, $K$-meshes of $10\times10\times10$ (ionic relaxations) and $20\times20\times20$ (self-consistent-field solutions) were used for PBE calculations. Cell volume, shape, and atomic positions for all the structures were fully relaxed using conjugate gradient algorithm until the energy (forces) converges to $10^{-6}$ eV (0.001 eV/Å). A tetrahedron method with Blochl corrections was used for accurate electronic density of states.



Density functional perturbation theory (DFPT) combined with phonopy\cite{phonopy} was used to obtain relevant phonon properties. Alloy Theoretic Automated Toolkit (ATAT)\cite{ATAT} was used to generate special quasi-random structures to simulate disordered phases of these compounds. Ab-initio Scattering and Transport (AMSET)\cite{AMSET} code was used to calculate the electronic transport properties which uses the variable carrier relaxation time to evaluate the transport distribution function while solving the Boltzmann transport equations. Debye-Callaway (DC) model\cite{dcmodel} was used to calculate lattice thermal conductivity. More details about this model is described in supplementary information.

\section{Results and Discussion}
Table \ref{Energies} shows the optimized lattice constant and the relative energies of cubic (F$\bar{4}$3m), tetragonal (I$\bar{4}$2d) and SQS structures (P1) of Hf$_2$Ni$_2$InSb and Zr$_2$Ni$_2$InSb. Six different configurations with different site occupancies of In and Sb were simulated using the conventional cell of cubic phase and the energetically most stable configuration is presented in the manuscript. 

\begin{table}[!h]
	\caption{Theoretically optimized lattice constants and relative energies of cubic, tetragonal and SQS structures of Hf$_2$Ni$_2$InSb and Zr$_2$Ni$_2$InSb} \label{Energies}
	
	\begin{tabular}{|c|c|c|c|c|}
		\hline
		  Compound  &Crystal  &Optimized lattice   & $\triangle$E (meV/  \\
	  &structure   &constants (\AA)     &atom)  \\  
		\hline

			 &cubic  &  a=b=c=6.11   &8.0  \\
			  
			Hf$_2$Ni$_2$InSb &tetragonal  &  a=b=6.11, c=12.24  &0.0      \\

			&SQS  &  a=6.09, b=6.12,c=5.94   &62.0   \\
			\hline
		
			  &cubic    &  a=b=c= 6.15  &6.0   \\
			  Zr$_2$Ni$_2$InSb &tetragonal   &  a=b=6.14, c=12.31   &0.0      \\
			
		&SQS    &  a=b=6.18, c=6.12  & 134.0    \\
		
		\hline
	\end{tabular}
\end{table}
Tetragonal structure of these compounds is theoretically predicted to be stable in open quantum materials database (OQMD). Solid solution phase (represented by SQS structure here) is usually inevitable for this class of compounds at high temperature.\cite{ordered}$^{,}$\cite{sqs} As such a comparative study of all these three phases is highly desired to facilitate an in-depth analysis of TE properties of these alloys in different T-range. Interestingly, the energy difference between the two ordered phases is very small (8 meV for Hf-based) and 6 meV for Zr-based double HH-alloys), where SQS phase is much higher in energy (60-130 meV) as compared to the lower energy tetragonal phase. 

\begin{figure}[h]
	\centering
	\includegraphics[scale=0.5]{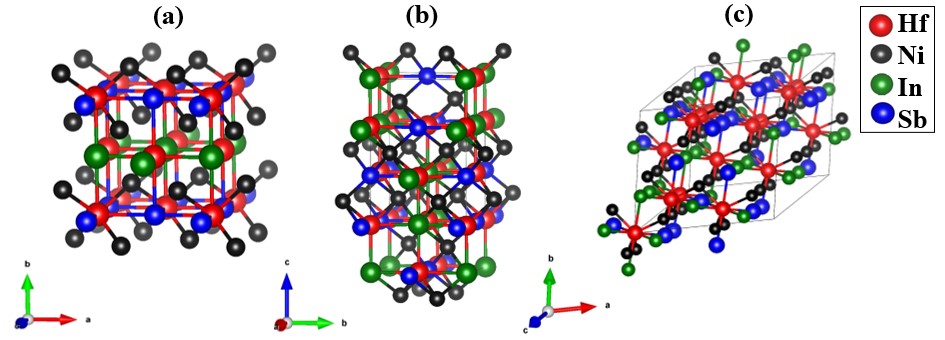}
	\includegraphics[scale=0.5]{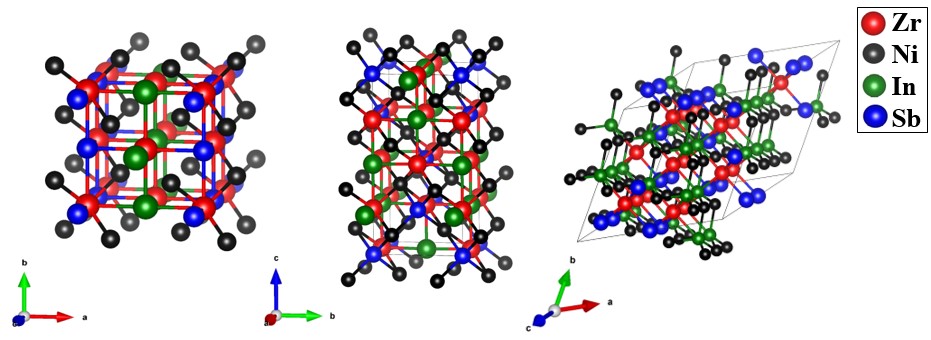}
	\caption{Theoretically optimized crystal structures of Hf$_2$Ni$_2$InSb (top) and Zr$_2$Ni$_2$InSb (bottom) in (a) cubic (b) tetragonal and (c) SQS phases.}
	\label{crystal}
\end{figure}

As evident from table \ref{Energies}, the SQS structure are off-cubic due to the presence of disorder.
Figure \ref{crystal} shows the theoretically optimized cubic, tetragonal and SQS structure of Hf$_2$Ni$_2$InSb (top) and Zr$_2$Ni$_2$InSb (bottom). Few bond lengths in both cubic and tetragonal structures are same i.e., d\textsubscript{Hf-In} = d\textsubscript{Hf-Sb} = 3.05 \AA, d\textsubscript{In-Ni} = 2.68 \AA, d\textsubscript{Sb-Ni} = 2.61 \AA (in Hf$_2$Ni$_2$InSb) and d\textsubscript{Zr-In} = d\textsubscript{Zr-Sb} = 3.07 \AA, d\textsubscript{In-Ni} = 2.69 \AA, d\textsubscript{Sb-Ni} = 2.63 \AA (in Zr$_2$Ni$_2$InSb), while  d\textsubscript{Hf-Ni} bond length in cubic and tetragonal structure are 2.61 \AA and 2.64 \AA respectively. The same for Zr$_2$Ni$_2$InSb are 2.63 and 2.66 \AA respectively. In SQS structure, there is large variation in bond length due to randomness, ranging from 2.6 to 6.04 \AA. 

\begin{figure}[!h]
	\centering
	\includegraphics[scale=0.6]{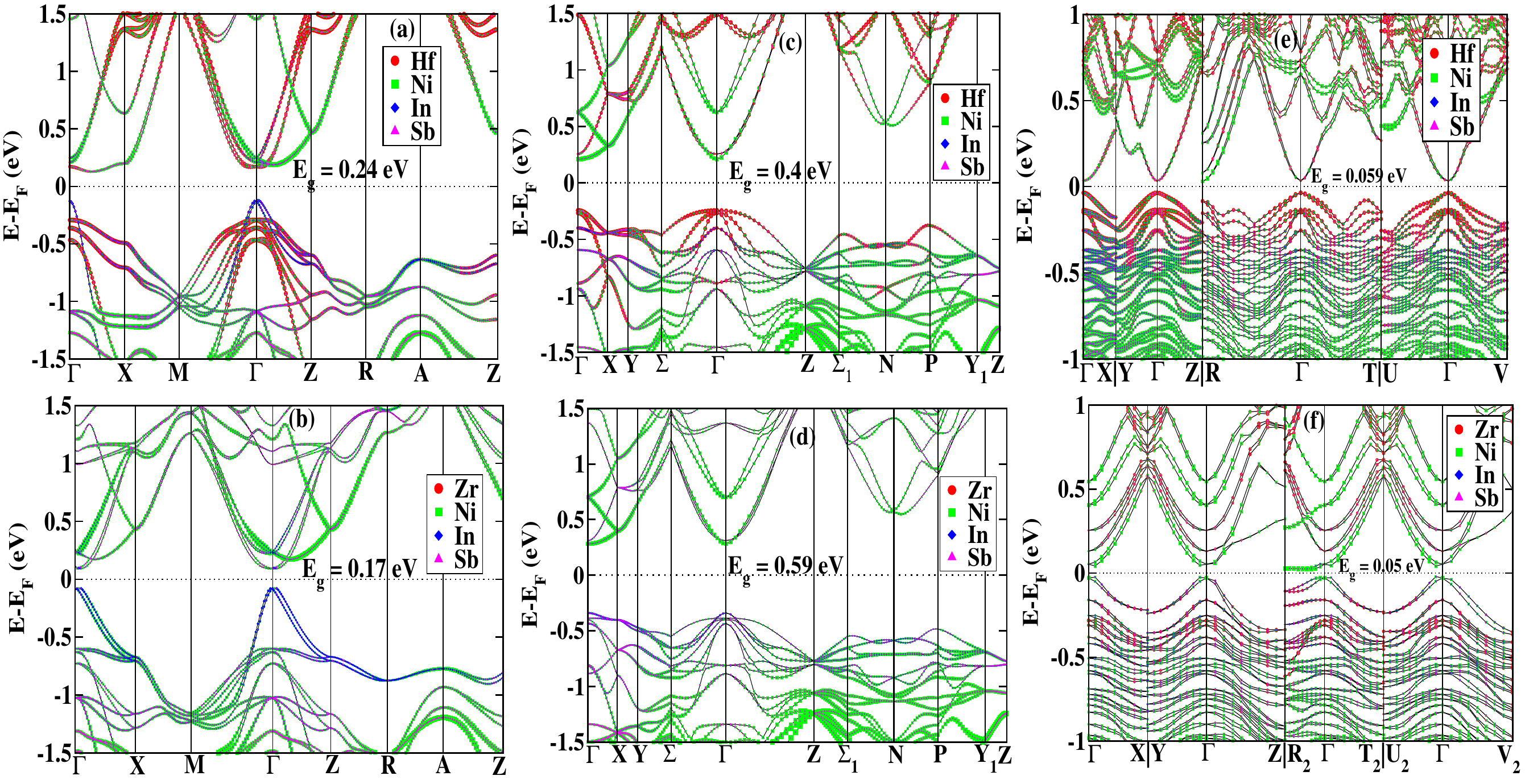}
	\caption{Atom/orbital-projected electronic band structures of (a,b) cubic (c,d) tetragonal and (e,f) disordered SQS structure of Hf- and Zr-based double HH alloys respectively.}
	\label{Hfbands}
\end{figure}
\subsection{Electronic structure}
Figures \ref{Hfbands}(a) and (b) show the atom-projected band structure of cubic Hf$_2$Ni$_2$InSb and Zr$_2$Ni$_2$InSb respectively. For Hf$_2$Ni$_2$InSb, the conduction bands are mostly contributed by Hf and Ni atoms whereas the valence band edges are composed of In and Sb atoms. Multiple valleys are favorable for thermoelectric performance of a compound as it leads to large band degeneracy. The second conduction band minima (CBM-2) and third conduction band minima (CBM-3) lie at an energy difference of 0.03 and 0.05 eV (also contributed by Hf and Ni atoms). For Zr$_2$Ni$_2$InSb, the conduction bands are mostly contributed by Ni atoms while valence bands by In atoms. The CBM-2 lies at an energy difference of 0.08 eV. This shows large conduction band degeneracy in the cubic phase of these compounds. The band gap calculated using HSE-SOC functional for Hf and Zr-based compounds are 0.24 eV and 0.17 eV respectively. The former (latter) is an indirect (direct) band gap semiconductor. SOC plays a crucial role due to heavy element Sb, and causes a reduction in the band gap as compared to non-SOC values.   

Figures \ref{Hfbands}(c) and (d) show the band structures of Hf$_2$Ni$_2$InSb and Zr$_2$Ni$_2$InSb respectively in their tetragonal phase. Both are direct band gap semiconductors with a value of 0.4 and 0.59 eV respectively. Valence band edges show dominant contribution from Hf and Ni atoms whereas conduction band edges are dominated by Ni atoms for Hf$_2$Ni$_2$InSb. For Zr$_2$Ni$_2$InSb, the dominant contribution arises from In and Ni atoms near valence band edges whereas conduction band edges are mostly composed of Ni. Figures \ref{Hfbands} (e) and (f) show the band structures of disordered phase for Hf and Zr-based compounds respectively. HSE-SOC band gap for Hf$_2$Ni$_2$InSb and Zr$_2$Ni$_2$InSb are 0.06 eV and 0.05 eV respectively. Hf-atoms have the dominant contribution near valence band edges while a mixed contribution from Hf and Ni-atoms near conduction band edges. Zr$_2$Ni$_2$InSb has a similar contribution with Hf replaced by Zr.

The ordered structures have larger band gaps as compared to the disordered SQS structures, which can help in suppressing the bipolar effect\cite{bipolar} and hence better electronic transport properties. The ordered tetragonal phase has the largest band gap with flat valence bands which leads to higher density of states effective mass and hence larger p-type thermopower Whereas conduction bands for ordered cubic phase are flat and show multiple valleys at very small energy difference. This indicates better n-type thermopower in ordered cubic phase. All the thermoelectric properties are calculated using PBE band structure with scissor shifted band gap evaluated from HSE-SOC functional.
 
\subsection{Electronic Transport}
Transport parameters of most thermoelectric materials are strongly dependent on carrier relaxation time ($\tau$). Most previous simulations are based on constant relaxation time approximation(CRTA),\cite{CRTA1}$^{,}$\cite{CRTA2} which is a very crude approximation. $\tau$ is dictated by different scattering mechanisms such as acoustic scattering, optical scattering, scattering by impurities and defects and electric polarization in case of the polar lattice. In the present work, we have estimated the relaxation time for electron and hole transport, using Ab-initio Scattering and Transport (AMSET)\cite{AMSET}. AMSET is a numerical code for calculating carrier relaxation time and transport properties within first principles framework. We have taken into account all four types of scattering mechanisms in our TE calculations i.e. acoustic deformation potential (ADP), ionized impurity scattering (IMP), polar optical phonon scattering (POP) and piezoelectric scattering (PIE). We have also captured the effect of charge carrier screening arising out of free carriers at high concentrations. The details of these mechanisms is given in supplementary information (SI).

The relaxation time ($\tau$) for Hf$_2$Ni$_2$InSb and Zr$_2$Ni$_2$InSb (in cubic phase) is shown in Figures \ref{tau_Hf} and \ref{tau_Zr} of supplementary information respectively. As can be seen, $\tau$ is strongly dependent on both carrier concentration (n) and temperature (T). For Hf$_2$Ni$_2$InSb, at a high temperature of 900 K, $\tau$ varies between 25-21 fs for holes and 35-20 fs for electrons with increasing carrier concentration. For Zr$_2$Ni$_2$InSb, $\tau$ varies between 26-22 fs for holes and 37-22 fs for electrons at 900 K.

\begin{figure}[!h]
	\centering
	\includegraphics[scale=0.8]{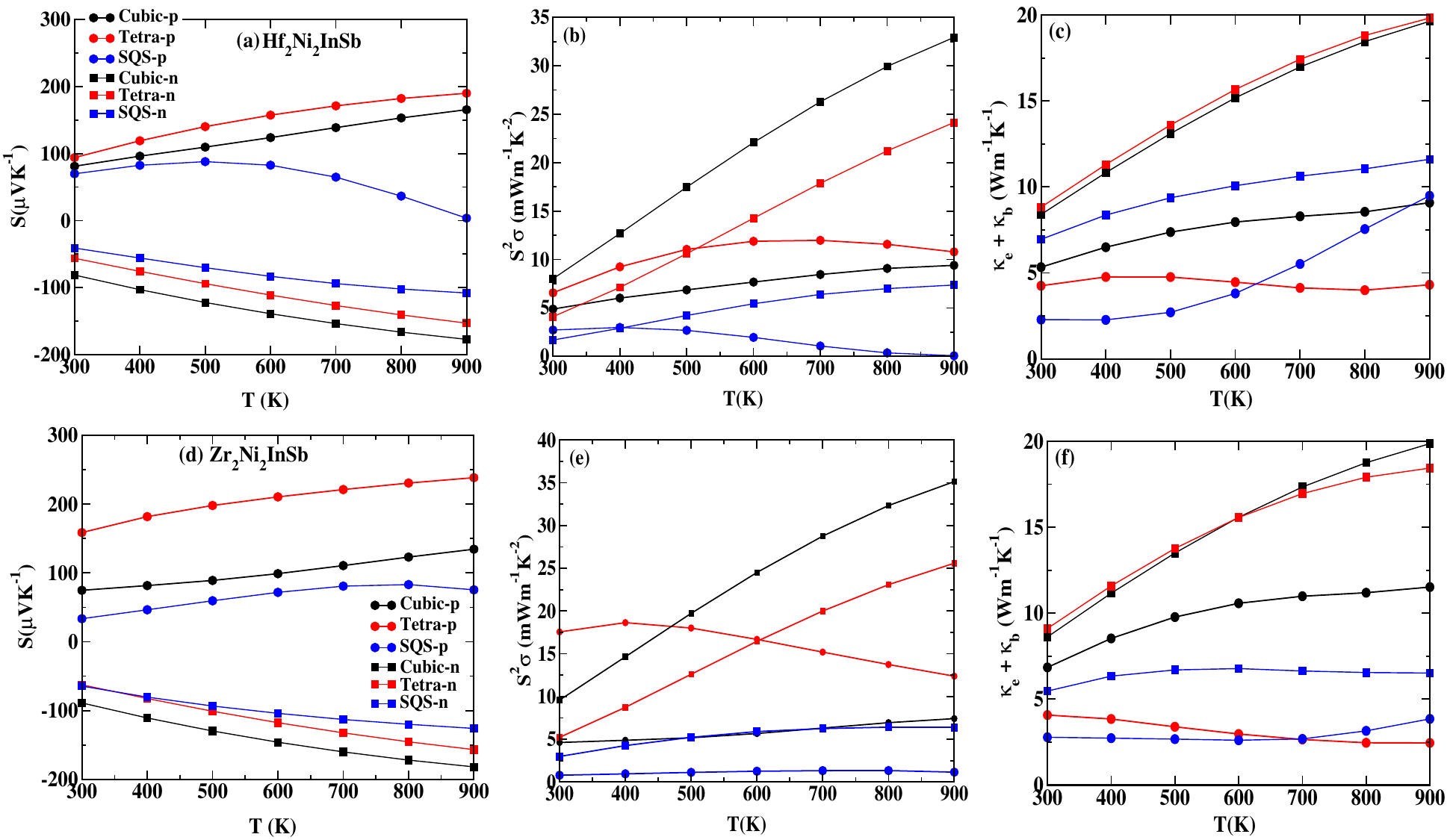}
\caption{Temperature dependence of (a,d) Seebeck coefficient(S), (b,e) power factor (S$^2$$\sigma$), (c,f) electronic thermal conductivity ($\kappa_e$+$\kappa_b$) for p-type (circle) and n-type (square) conduction in cubic (black line), tetragonal (red line) and SQS (blue line) structures of Hf$_2$Ni$_2$InSb and Zr$_2$Ni$_2$InSb respectively at a fixed  carrier concentration of 1 $\times$ 10$^{21}$ cm$^{-3}$.}
	\label{Hfcubic_dHH}
\end{figure}
\begin{figure*}[t]
	\centering
	\includegraphics[scale=0.55]{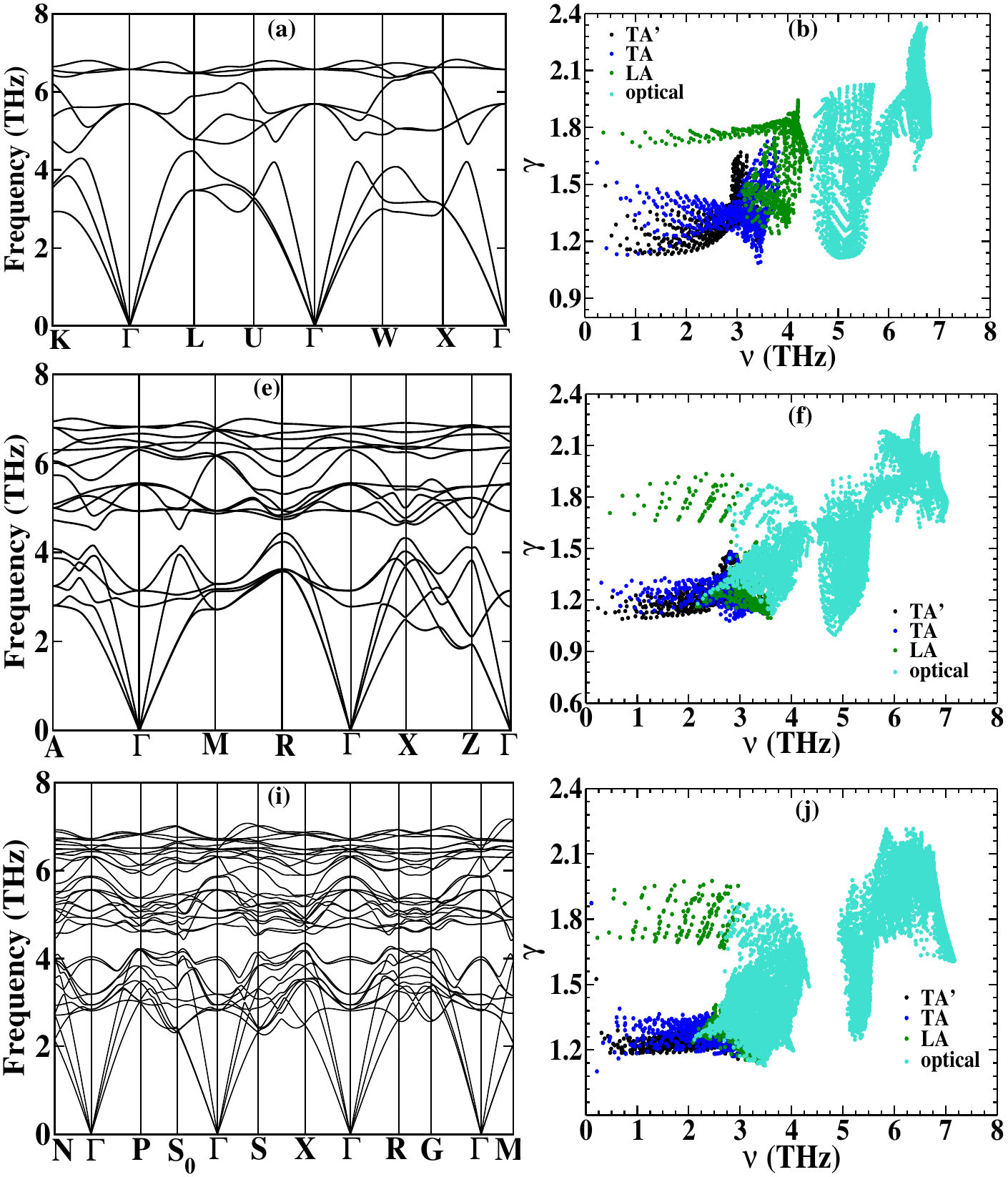}\includegraphics[scale=0.55]{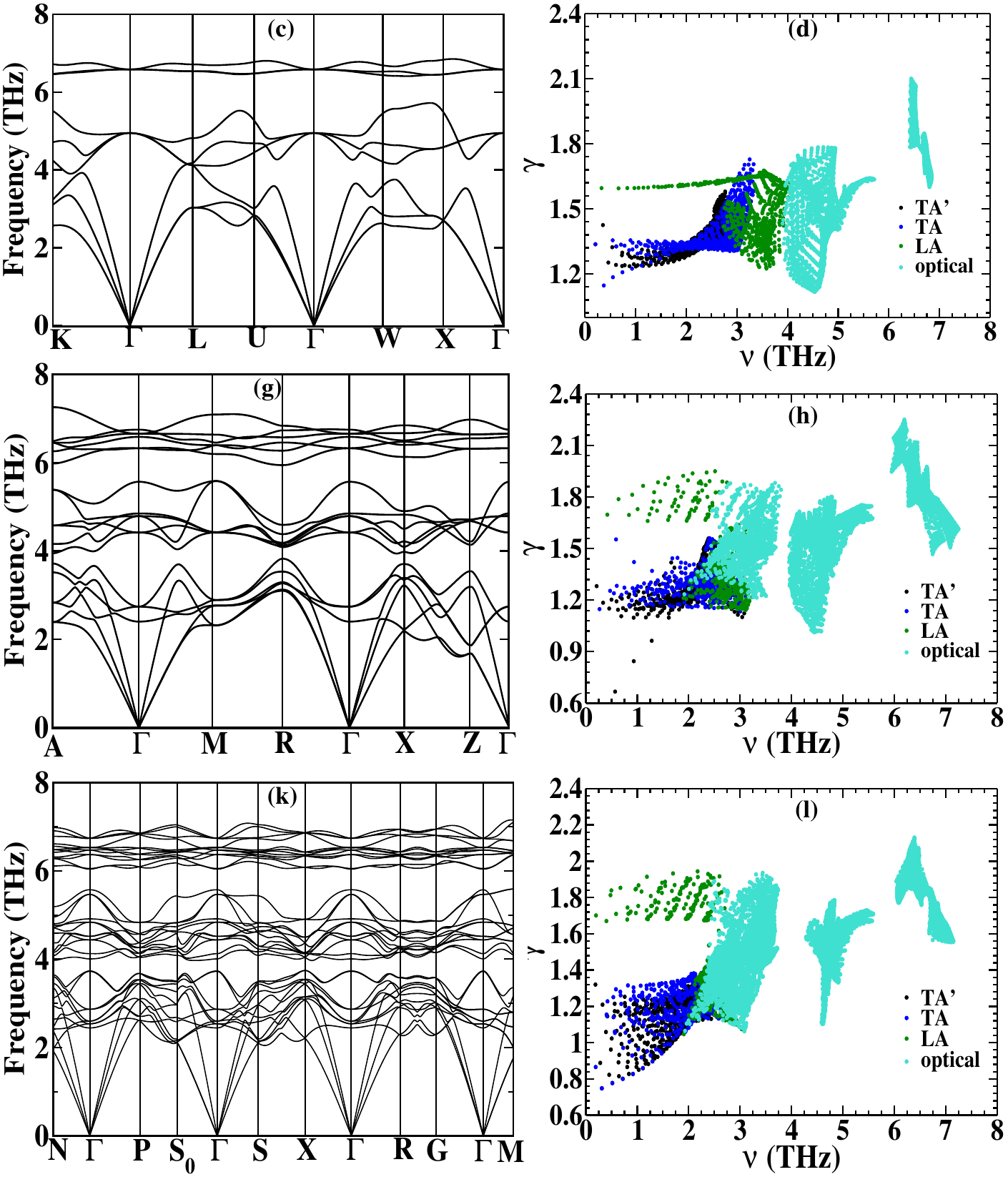}
\caption{Phonon dispersion and mode Gruneisen parameter for ternary (a,b) HfNiSn and (c,d) ZrNiSn; cubic double HH (e,f) Hf$_2$Ni$_2$InSb and (g,h) Zr$_2$Ni$_2$InSb; tetragonal double HH (i,j) Hf$_2$Ni$_2$InSb and (k,l) Zr$_2$Ni$_2$InSb respectively.}
	\label{HfPhonongrun}
\end{figure*}
 
\begin{figure}[!htbp]
	\centering
	\includegraphics[scale=0.7]{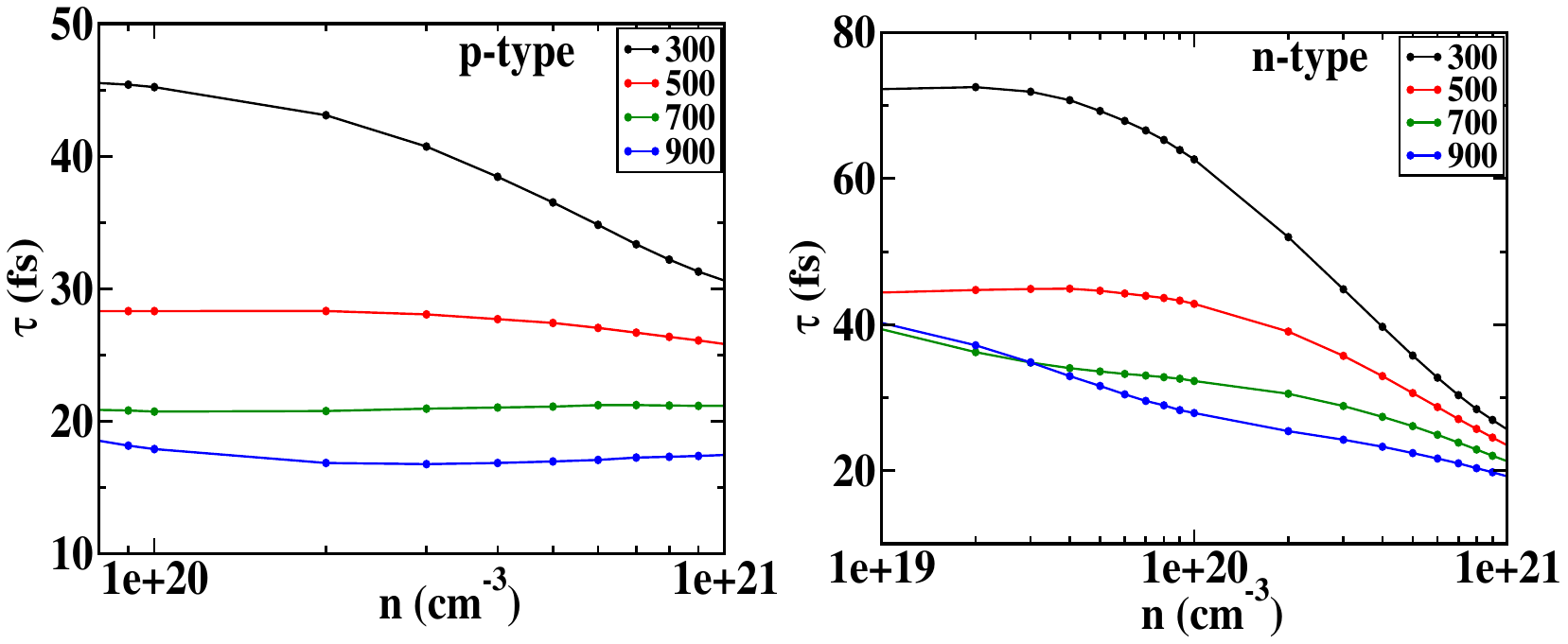}
	\caption{Temperature dependence of p-type (holes) and n-type(electrons) relaxation time for tetragonal Hf$_2$Ni$_2$InSb.}
	\label{tau_Hf_T}
\end{figure}

\begin{figure}[!h]
	\centering
	\includegraphics[scale=0.75]{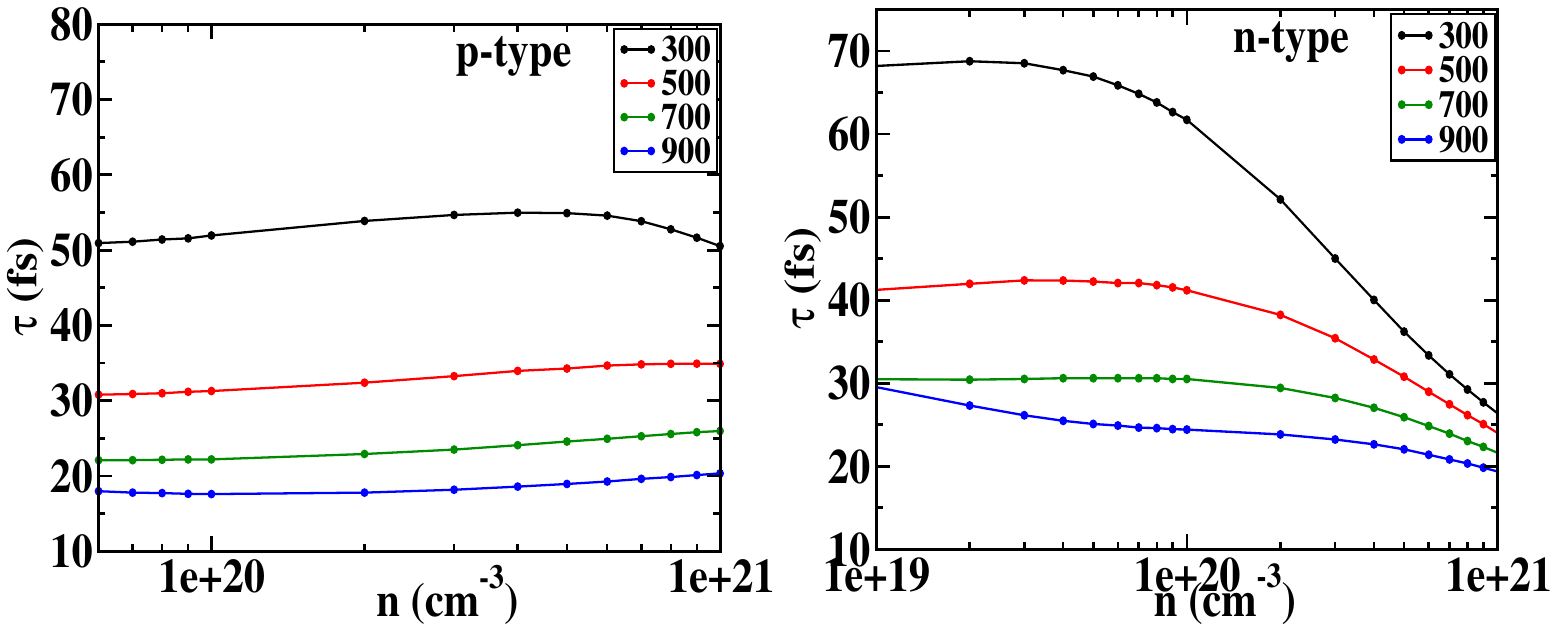}
	\caption{Temperature dependence of p-type (holes) and n-type(electrons) relaxation time for tetragonal Zr$_2$Ni$_2$InSb.}
	\label{tau_Zr_T}
\end{figure}

The relaxation time ($\tau$) of the two compounds in tetragonal phase are shown in Figure \ref{tau_Hf_T} and \ref{tau_Zr_T} respectively. For  Hf$_2$Ni$_2$InSb, $\tau$ varies between 18-17 fs for holes and 40-19 fs for electrons at 900 K. For Zr$_2$Ni$_2$InSb, $\tau$ varies between 18-20 fs for holes and 30-20 fs for electrons at 900 K. Thus, the order of magnitude of $\tau$ do not vary much, as we go from one phase to the other for a given compound. For SQS structure, we expect a similar or relatively lower value of $\tau$. Figure \ref{Hfcubic_dHH} shows a comparison of the T-dependent electronic transport properties of Hf and Zr-based compounds in different phases at a fixed carrier concentration of 1 $\times$ 10$^{21}$ cm$^{-3}$. The choice of carrier concentration is guided by a previous experimental report on Zr-based double HH-alloy\cite{ZrNIInSb}. The ordered structures show better electronic properties than the SQS structures. Also, as expected from electronic band structures topology, the p-type thermopower of tetragonal structures is larger whereas for cubic phase, n-type thermopower is most promising for both the compounds (see figs. \ref{Hfcubic_dHH}(a,d)). As a result, power-factor of p-type tetragonal phase and n-type cubic phase is largest in a large T-range. (see figs. \ref{Hfcubic_dHH}(b,e)). Figures \ref{Hfcubic_dHH}(c,f) shows the T-dependence of thermal conductivity ($\kappa_e$+$\kappa_b$) for both the compounds, where $\kappa_b$ is bipolar thermal conductivity. The ($\kappa_e$+$\kappa_b$) values of SQS structures show a rise at higher T (around 500-600 K) because of the bipolar component of $\kappa$. This is also the reason for a slight decrease in Seebeck coefficient at higher T for SQS structures. The ordered phases show larger and comparable values of electronic thermal conductivity for n-type. 

These double half heusler compounds show promising electronic transport properties (large power-factor) due to favorable band features such as flat bands in the tetragonal phase and high band degeneracy in cubic phase. Thus, the present compounds supersede/compete well with previously reported high-thermoelectric performance materials. For example, ZrCoBi$_{0.65}$Sb$_{0.15}$Sn$_{0.20}$ shows a ZT value of 1.42 at around 970 K.\cite{HH1} The power-factor for this compound is around 3.8 mWm$^{-1}$K$^{-2}$. Another promising p-type HH compound FeNb$_{0.8}$Ti$_{0.2}$Sb shows a ZT value of 1.1 (at around 970 K)\cite{HH2} with the corresponding   power factor of 5.3 mWm$^{-1}$K$^{-2}$. For n-type HH compounds,
Ti$_{0.5}$Zr$_{0.25}$Hf$_{0.25}$NiSn$_{0.998}$Sb$_{0.002}$ shows a ZT value of 1.5 (at 700 K)\cite{HH3} and a power-factor of 6.2 mWm$^{-1}$K$^{-2}$. Hf$_{0.6}$Zr$_{0.4}$NiSn$_{0.995}$Sb$_{0.005}$ is yet another system with a promising ZT value of 1.2 (at 900 K)\cite{HH4} and a power-factor of 4.7 mWm$^{-1}$K$^{-2}$. In double half heusler family, Ti$_4$Fe$_2$Ni$_2$Sb$_4$(in solid solution) shows a ZT value of 0.5 and 0.4 for p-type and n-type conduction with power-factor values of around 1.7 and 1.2  mWm$^{-1}$K$^{-2}$ respectively at around 900 K.\cite{Liu} At the same temperature, the ordered structure for the same compound shows ZT value of 1.5 for p-type and 0.5 for n-type.\cite{ordered} The corresponding theoretically calculated power factor values are around 7 mWm$^{-1}$K$^{-2}$ and 4 mWm$^{-1}$K$^{-2}$.
Hf$_2$Ni$_2$InSb and Zr$_2$Ni$_2$InSb (see Fig \ref{Hfcubic_dHH}) show much higher power-factor values at similar temperature range. Hf$_2$Ni$_2$InSb has a value of  9.2  (10.7) mWm$^{-1}$K$^{-2}$ and 32 (24) mWm$^{-1}$K$^{-2}$ for for p-type and n-type cubic (tetragonal) structure respectively, while Zr$_2$Ni$_2$InSb has a value of 7.4 (12) mWm$^{-1}$K$^{-2}$ and 35 (25.4) mWm$^{-1}$K$^{-2}$ for p-type and n-type cubic (tetragonal) structures respectively. The disordered (SQS) phases also show comparable/better power-factor values (7.3 mWm$^{-1}$K$^{-2}$ for Hf-based and 6.4 mWm$^{-1}$K$^{-2}$ for Zr-based alloy).
	
\subsection{Phonon Transport}
 For comparison sake, we have not only calculated the phonon properties (specially the order of magnitude of lattice thermal conductivity, $\kappa_L$) of the present double HH alloys, but also their ternary counterparts ZrNiSn and HfNiSn (with same VEC) which are extremely studied in the literature and reported to crystallize in F$\bar{4}$3m ($\sharp$216) structure.\cite{ZrNiSn,APL79,APL86} 
 Figures \ref{HfPhonongrun}(a,c), \ref{HfPhonongrun}(e,g) and \ref{HfPhonongrun}(i,k) show the phonon dispersion for ternary (HfNiSn, ZrNiSn), cubic double HH (Hf$_2$Ni$_2$InSb, Zr$_2$Ni$_2$InSb) and tetragonal double HH (Hf$_2$Ni$_2$InSb, Zr$_2$Ni$_2$InSb) alloys respectively. The ternary, cubic and tetragonal phases have 3-atoms, 6-atoms and 12-atoms in the primitive unit cell giving rise to 9, 18 and 36 phonon branches respectively. Of them, the lowest 3 branches are acoustic and the rest corresponds to optical branches respectively. 
 \begin{figure}
	\centering
	\includegraphics[scale=0.8]{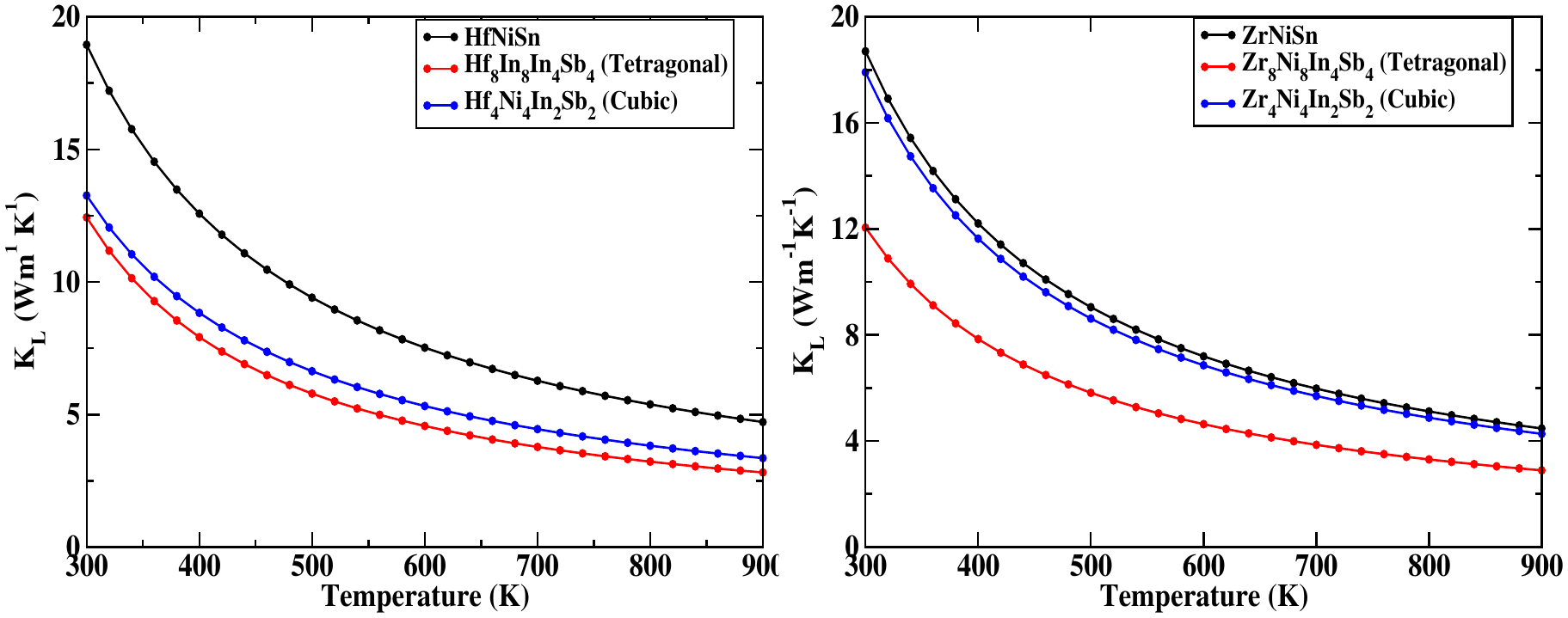}
	\caption{Comparison of simulated lattice thermal conductivity ($\kappa_L$) of ternary XNiSn with those of cubic and tetragonal phases of double HH X$_2$Ni$_2$InSb for (left) Hf-based and (right) Zr-based compounds.}
	\label{kL_Hf}
\end{figure} 
 The acoustic branches are further classified into one longitudinal (LA) and two transverse (TA, TA$^{'}$) modes. The velocity of each acoustic mode ($\gamma$) is calculated from the slope of the band corresponding to the vibrational band at $\Gamma$-point. Debye temperature ($\theta$) can be estimated from the maximum frequency corresponding to the given vibrational mode, within a reasonable approximation. Apart from $\nu$ and $\theta$, DC model for $\kappa_L$ also requires few other quantities including gruneisen parameter for HfNiSn, cubic Hf$_2$Ni$_2$InSb and tetragonal Hf$_2$Ni$_2$InSb respectively. The same for Zr-based compounds are shown in Figs. \ref{HfPhonongrun}(d), \ref{HfPhonongrun}(h) and \ref{HfPhonongrun}(l). The simulated values of various parameters, phonon group velocities ($\nu_i$), Debye temperature ($\theta_i$), cell volume (V), atomic mass (M), Gruneisen parameter ($\gamma_i$ for different vibrational modes (LA, TA, TA$^{'}$) for the three Hf and Zr-based compounds are presented in SI.
Larger values of $\gamma$ indicates high degree of anharmonicity. The group velocity of acoustic phonons is reduced in double half heuslers (cubic phase) as compared to their ternary counterparts due to large mixing of acoustic and optical phonon modes in the former. This indicates reduced lattice thermal conductivity for double HH as compared to ternary alloys.

\subsection{Thermal transport properties}

Figures \ref{kL_Hf} show the comparison of $\kappa_L$ for ordered phases of double half heuslers and their corresponding ternary counterparts. The simulated values of $\kappa_L$ vary between 13.3 to 3.3 Wm$^{-1}$K$^{-1}$ for Hf$_2$Ni$_2$InSb (in cubic phase) and 12.4 to 2.8  Wm$^{-1}$K$^{-1}$ (in tetragonal phase) whereas for ternary HfNiSn, it ranges between 18.9 to 4.7 Wm$^{-1}$K$^{-1}$ in the temperature range 300-900 K. For Zr$_2$Ni$_2$InSb, $\kappa_L$ varies between 17.8 to 4.3 Wm$^{-1}$K$^{-1}$ (for cubic) and 12.1 to 2.9 Wm$^{-1}$K$^{-1}$ (for tetragonal) in the temperature range 300-900 K whereas it varies between 18.7 to 4.5 Wm$^{-1}$K$^{-1}$ for ZrNiSn for the same temperature range. The previous theoretically reported room temperature $\kappa_L$ values of ZrNiSn and HfNiSn are 19.69 and 18.5 Wm$^{-1}$K$^{-1}$, whereas for Zr$_2$Ni$_2$InSb and  Hf$_2$Ni$_2$InSb (in tetragonal phase), the values are  13.58  and 12.5 Wm$^{-1}$K$^{-1}$ respectively at 300 K.\cite{double_HH1} These values are in fair agreement with our simulated values obtained from DC model.
Clearly, we can see the reduction of lattice thermal conductivity in double half heuslers as compared to the corresponding ternary alloys which is definitely useful for enhancing the TE figure of merit (ZT). For disorder SQS phase, we expect a further reduction of $\kappa_L$ due to enhanced disorder induced scatterings.
 
\section{Thermoelectric performance}
\begin{figure}[!h]
	\centering
	\includegraphics[scale=0.5]{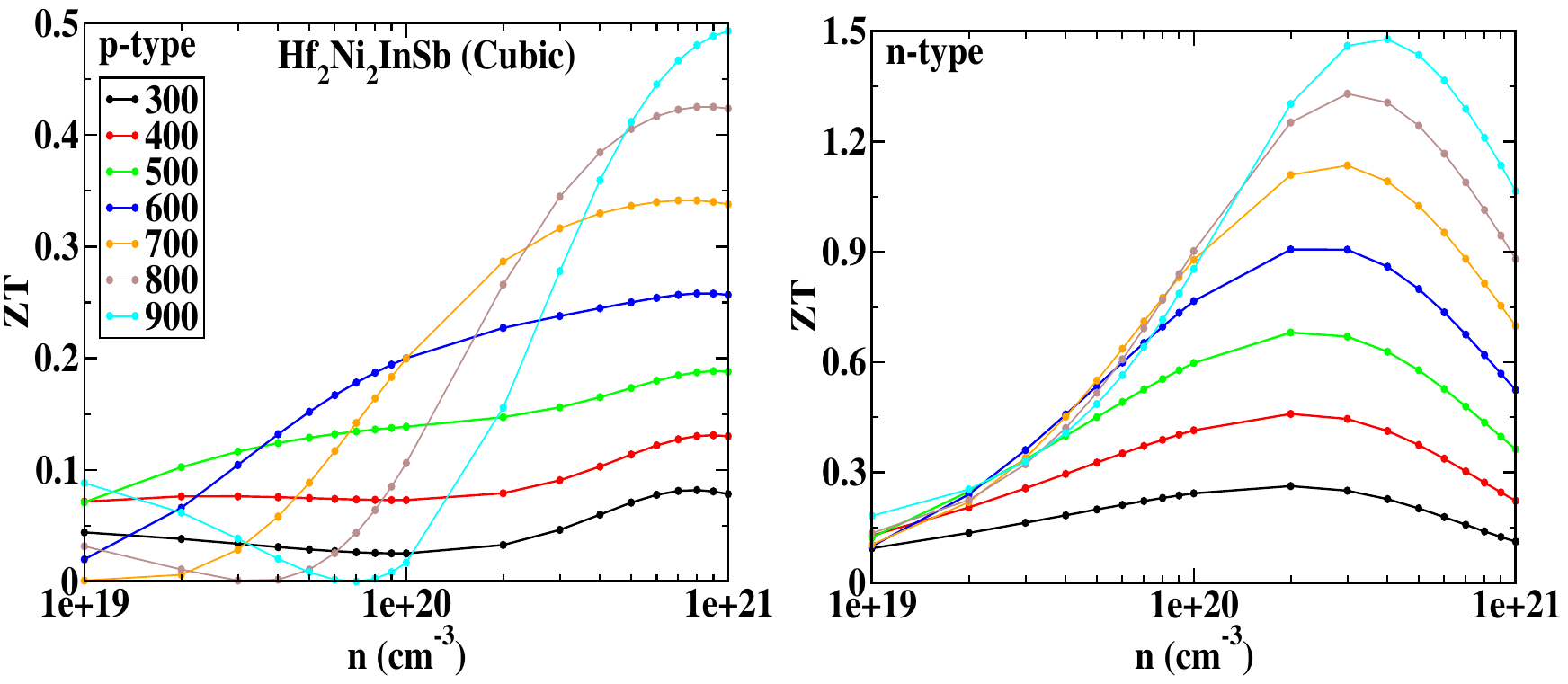}
	\includegraphics[scale=0.5]{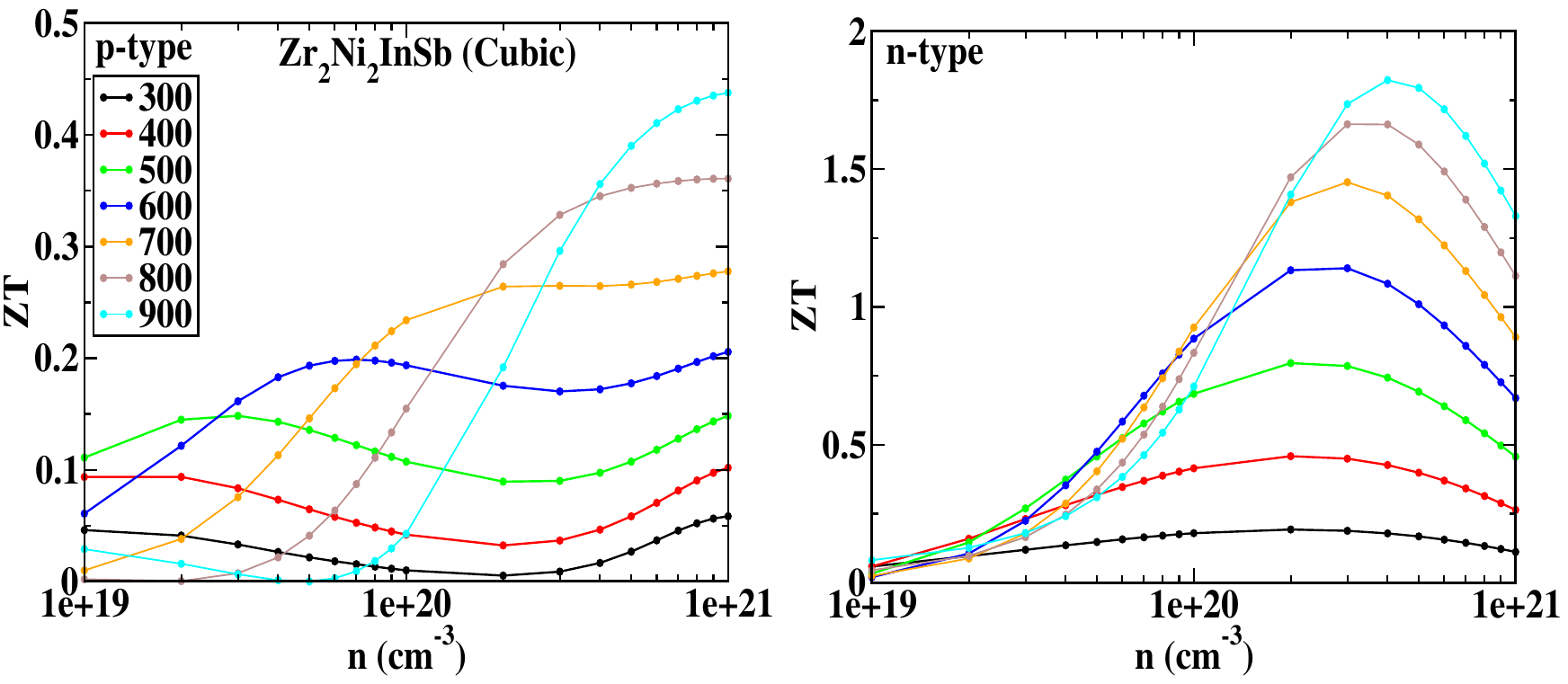}
	\caption{Thermoelectric figure of merit (ZT) as a function of carrier concentration (n) at different temperatures for cubic Hf$_2$Ni$_2$InSb (above) and Zr$_2$Ni$_2$InSb (below) alloy for p-type (left) and n-type (right) conduction. }
	\label{Zr_ZT}
\end{figure}

\begin{figure}[!h]
	\centering
	\includegraphics[scale=0.5]{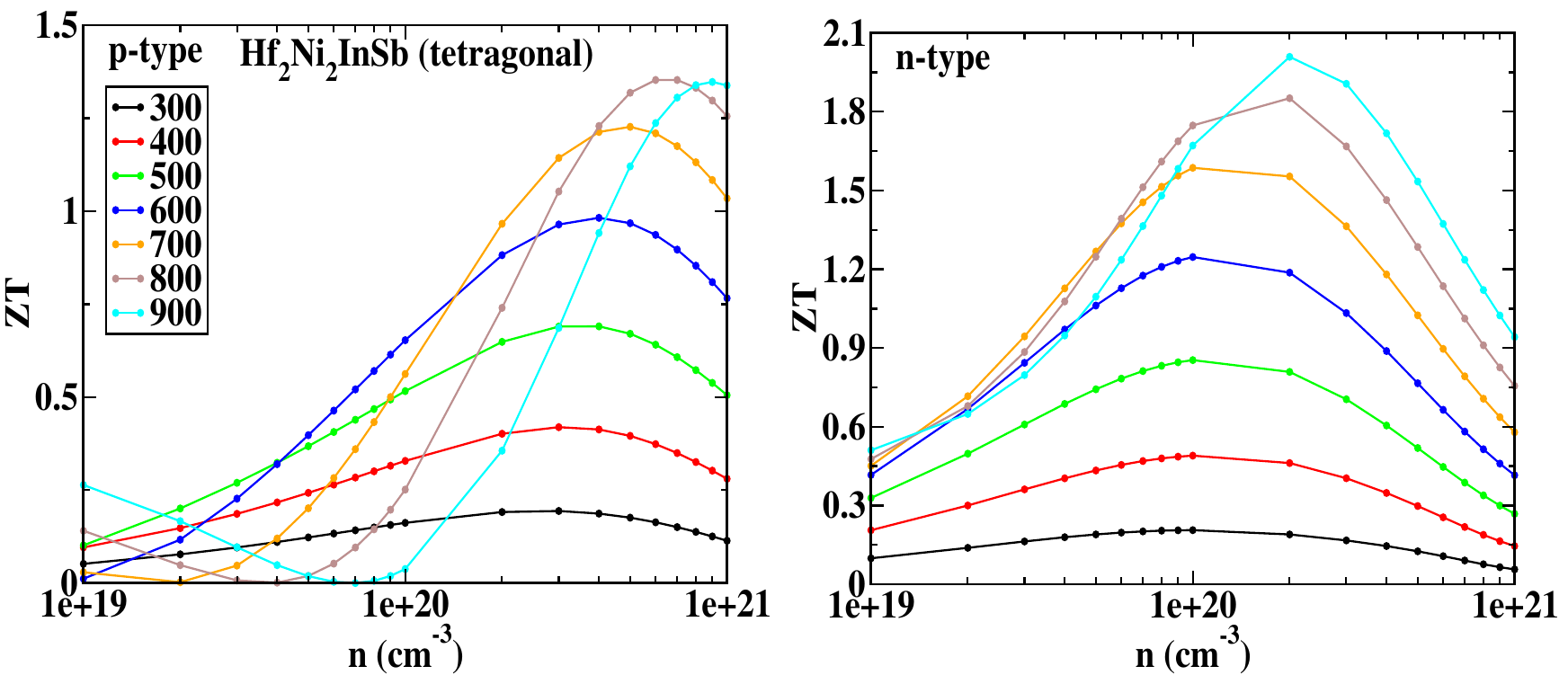}
	\includegraphics[scale=0.5]{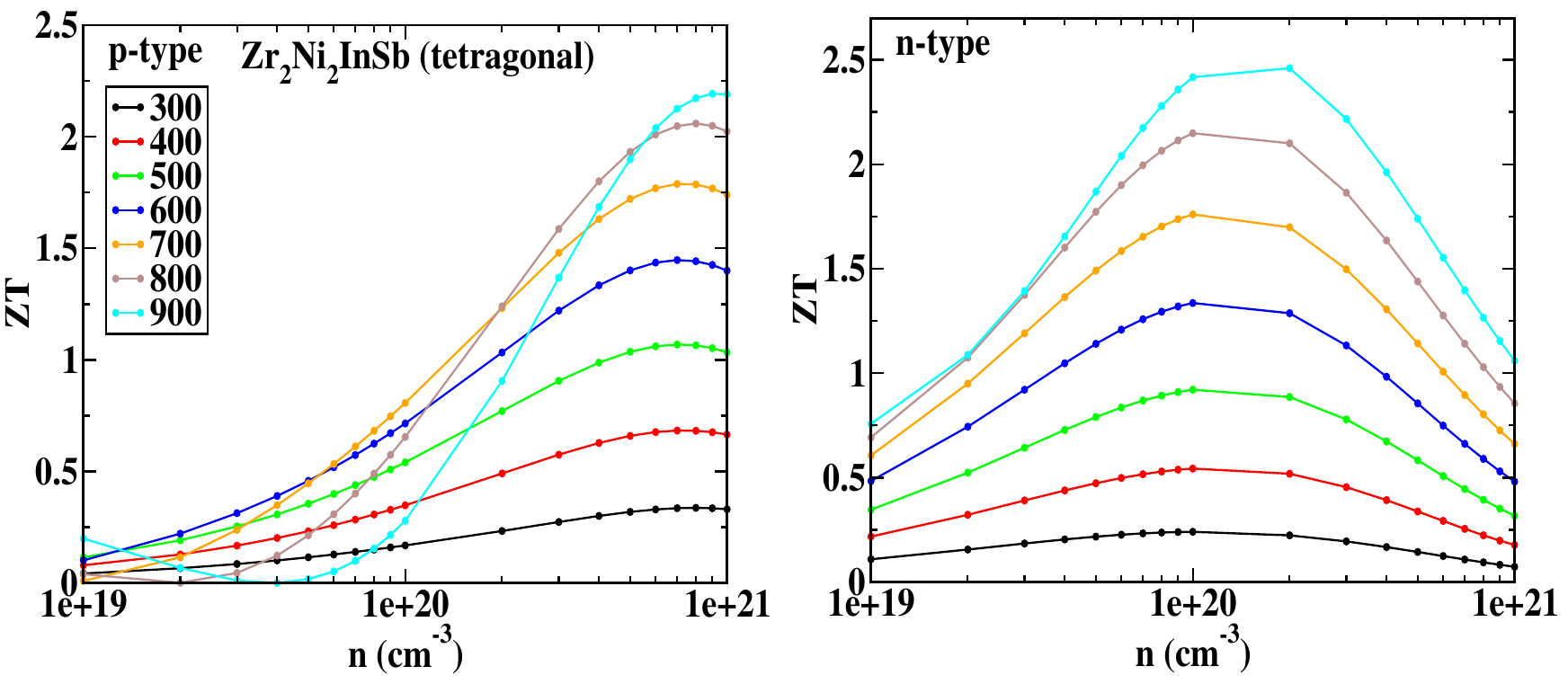}
	\caption{The thermoelectric figure of merit (ZT) as a function of carrier concentration (n) at different temperatures (T) for tetragonal Hf$_2$Ni$_2$InSb (above) and Zr$_2$Ni$_2$InSb (below) alloy for p-type (left) and n-type (right) conduction. }
	\label{Hf_ZT}
\end{figure}

As evident from Fig. \ref{Hfcubic_dHH}, the power factor for ordered phases is reasonably high (better or comparable to some of the best TE materials in the literature,\cite{HH1}$^{,}$\cite{HH4} along with the relatively low lattice thermal conductivity (see Fig. \ref{kL_Hf}). This indicates their potential for efficient TE performance. In this section, we shall focus on a comparison of TE figure of merit (ZT) for n and p-type conduction of ordered phases of these two alloys. Figure \ref{Zr_ZT} and \ref{Hf_ZT} show the carrier concentration dependence of ZT at different T for cubic and tetragonal phases of the two alloys respectively. The left (right) panel indicates the result for p-type (n-type) conduction. The change in carrier concentration (n) can be thought of as mimicking the effect of doping/alloying the host material, keeping the topology of band structure intact (the so-called rigid band approximation). 

For cubic phase,  Hf$_2$Ni$_2$InSb show a peak ZT value of 0.49 for p-type conduction at a carrier concentration of 1 $\times$ 10$^{21}$ cm$^{-3}$ and 1.48 for n-type at a carrier concentration of 4 $\times$ 10$^{20}$ cm$^{-3}$ respectively at 900 K. At these carrier concentrations, the maximum value of simulated Seebeck coefficient (S$_{max}$) and power factor (PF$_{max}$) are 165.5 $\mu$VK$^{-1}$ and 9.39 mWm$^{-1}$K$^{-2}$ for p-type and 248.9 $\mu$VK$^{-1}$ and 29.6 mWm$^{-1}$K$^{-2}$ for n-type conduction respectively. The maximum ZT value for Zr$_2$Ni$_2$InSb is 0.44 at a carrier concentration of 1 $\times$ 10$^{21}$ cm$^{-3}$ for p-type and 1.82 at a carrier concentration of 4 $\times$ 10$^{20}$ cm$^{-3}$ for n-type respectively at 900 K. The simulated S$_{max}$ and PF$_{max}$ at these carrier concentrations are 134.6 $\mu$VK$^{-1}$ and 7.41 mWm$^{-1}$K$^{-2}$ for p-type and are 250.4 mWm$^{-1}$K$^{-2}$ and 29.7 mWm$^{-1}$K$^{-2}$ for n-type respectively.

The tetragonal phase of both alloys shows the lowest value of lattice thermal conductivity along with better p-type electronic transport properties. 
The peak value of ZT for Hf$_2$Ni$_2$InSb for p-type and n-type conduction occur at carrier concentrations of 7 $\times$ 10$^{20}$ cm$^{-3}$ and 2 $\times$ 10$^{20}$ cm$^{-3}$ respectively. At these values of carrier concentrations, the  S$_{max}$ and PF$_{max}$ for p-type are 210.6 $\mu$VK$^{-1}$ and 10.96 mWm$^{-1}$K$^{-2}$, and for n-type are 270.1 $\mu$VK$^{-1}$ and 19.48 mWm$^{-1}$K$^{-2}$ respectively.
This gives a ZT value of $\sim$ 1.35 at 800 K for p-type and $\sim$ 2.0 at 900 K for n-type conduction.
For Zr$_2$Ni$_2$InSb, the peak value of ZT for p-type and n-type conduction is obtained at carrier concentration of  9 $\times$ 10$^{20}$ cm$^{-3}$ and  2 $\times$ 10$^{20}$ cm$^{-3}$ respectively. The S$_{max}$ and PF$_{max}$ for p-type are 247.2 $\mu$VK$^{-1}$ and 11.96 mWm$^{-1}$K$^{-2}$ and for n-type are 282.5 $\mu$VK$^{-1}$ and 20.03  mWm$^{-1}$K$^{-2}$ respectively. As expected, a high ZT value of  $\sim$ 2.19 and $\sim$ 2.46 at 900 K for p-type and n-type conduction were obtained for tetragonal Zr$_2$Ni$_2$InSb. 

Although the power factor of these double half heuslers is quite high in cubic phase for n-type conduction, the alloys actually show higher ZT value in tetragonal phase due to lower values of lattice thermal conductivity. Thus, the reduction of lattice thermal conductivity along with enhanced power factor leads to the improvement of thermoelectric performance for these double half-heusler compounds as compared to corresponding 18 VEC ternary half-heusler alloys.

\section{Conclusion}
In summary, we report an ab-initio study of two double half heusler alloys, X$_2$Ni$_2$InSb (X=Hf/Zr), in their three competing structural phases: tetragonal, cubic and solid solution. Double half-heusler alloys are formed via the transmutation of two single heusler compounds and hence have higher flexibility for tuning their properties. The simulated band gaps (using HSE06 hybrid functional) for Hf$_2$Ni$_2$InSb lie in the range 0.06-0.4 eV while those for Zr$_2$Ni$_2$InSb in 0.05-0.59 eV depending on the structure. Spin-orbit coupling plays a crucial role in splitting the bands due to heavy Sb-atom. Thermoelectric performance is mostly dominated by the promising electronic transport in these alloys, with ZT value as high as 2.46 for n-type Zr$_2$Ni$_2$InSb and 2.0 for n-type Hf$_2$Ni$_2$InSb at high temperature. Tetragonal phase show a relatively lower lattice thermal conductivity, responsible for the best thermoelectric figure of merit (ZT). The TE performance of p-type conduction is almost equally promising in tetragonal phase, with highest ZT of $\sim$ 1.35 and $\sim$ 2.19 for Hf- and Zr-based compounds. We expect the present study to receive immediate attention from experimentalists with a possibility to synthesize and cross validate our findings.



\section*{Supporting Information}
Here, we present auxiliary information on the formalism of lattice thermal conductivity and carrier relaxation time calculation.
\subsection{Lattice Thermal Conductivity}
Lattice thermal conductivity ($\kappa_L$) is calculated using the Debye-Callaway (D-C) model\cite{dcmodel}. The Debye-Callaway (DC) model has been proven to be useful  for estimating the lattice thermal conductivity for various experimentally synthesized compounds.\cite{JMCA}$^{,}$\cite{Ourpaper}$^{,}$\cite{our_JPCL} Comparison of $\kappa_L$ for three systems (Cu3SbSe4, Cu3SbSe3 and SnSe) with the previous reported experimental values have been shown in  reference\cite{dcmodel}.
Within this model, the major contribution to the lattice thermal conductivity comes from the 3-phonon scattering process  (normal $\&$ Umklapp phonon scattering) and is majorly due to the 3 acoustic modes of vibrations. The thermal contribution from each of the acoustic vibrational mode in D-C model\cite{dcmodel,Ourpaper} is given as,     
\begin{equation}
\kappa_{L}^{i}=C_{i}T^{3}/3\left\{ \intop_{0}^{\theta_{i}/T}\frac{\tau_{c}^{i}(x)x^{4}e^{x}}{\left(e^{x}-1\right)^{2}}dx+\frac{\left[\intop_{0}^{\theta_{i}/T}\frac{\tau_{c}^{i}(x)x^{4}e^{x}}{\tau_{N}^{i}\left(e^{x}-1\right)^{2}}dx\right]^{2}}{\intop_{0}^{\theta_{i}/T}\frac{\tau_{c}^{i}(x)x^{4}e^{x}}{\tau_{N}^{i}\tau_{U}^{i}\left(e^{x}-1\right)^{2}}dx}\right\}
\end{equation}

\noindent where $x=(\hbar\omega/k_B T)$ and the index $i$ denotes LA, TA and TA$^\prime$ modes of vibration. The constant $C_i$ is given by, 
\begin{equation}
C_{i}=\frac{k_{B}^{4}}{2\pi^{2}\hbar^{3}\nu_{i}}
\end{equation}

The scattering rates corresponding to the normal and Umklapp scattering processes are,
\begin{equation}
\frac{1}{\tau_{N}^{LA}(x)}=\frac{k_{B}^{3}\gamma_{LA}^{2}V}{M\hbar^{2}\nu_{LA}^{5}}\left(\frac{k_{B}}{\hbar}\right)^{2}x^{2}T^{5}
\end{equation}

\begin{equation}
\frac{1}{\tau_{N}^{TA/TA'}(x)}=\frac{k_{B}^{4}\gamma_{TA/TA'}^{2}V}{M\hbar^{3}\nu_{TA/TA'}^{5}}\left(\frac{k_{B}}{\hbar}\right)xT^{5}
\end{equation}

\begin{equation}
\frac{1}{\tau_{U}^{i}(x)}=\frac{\hbar\gamma^{2}}{M\nu_{i}^{2}\theta_{i}}\left(\frac{k_{B}}{\hbar}\right)^{2}x^{2}T^{3}e^{-\theta_{i}/3T}
\end{equation}

\noindent where, $k_B$ is the Boltzmann's constant, $\hbar$ is the reduced Plank's constant, $T$ is the temperature, $\theta$ is the Debye temperature, $V$ and $M$ are the volume and mass per atom respectively, $\gamma$ is the mode greenness parameter, $\nu$ is phonon group velocity and $\omega$ is the angular frequency.
The input parameters such as the mode velocities, Debye temperature, Gruneisen parameter, etc were calculated from the phonon band structure and the mode Gruneisen parameter from the Phonopy code,\cite{phonopy}
which was used as post processing tool after doing the density functional perturbation theory (DFPT) calculations in Vienna Ab-initio Simulation Package (VASP).

\subsection{Simulated parameters needed for $\kappa_L$ calculations}
The numerical values of various parameters used to calculate ($\kappa_L$) for HfNiSn are; $\nu_{LA}$ = 4746.3246 m/s, $\nu_{TA}$ = 3120.5535 m/s, $\nu_{TA^{'}}$ = 2508.8948 m/s, $\gamma_{LA}$ = 1.55, $\gamma_{TA}$ = 1.38,$\gamma_{TA^{'}}$ = 1.35, $\gamma$= 1.59, $\theta_{LA}$ = 195.65 K, $\theta_{TA}$ = 160.06 K, $\theta_{TA^{'}}$ = 145.33 K, V = 19.05 $\times$ $10^{-30}$ $m^{3}$, and M = 196.93 $\times$ $10^{-27}$ kg. 
The numerical values of various parameters used to calculate ($\kappa_L$) for Hf$_2$Ni$_2$InSb (cubic phase) are; $\nu_{LA}$ = 4152.7021 m/s, $\nu_{TA}$ = 3438.3485 m/s, $\nu_{TA^{'}}$ = 2304.3666 m/s, $\gamma_{LA}$ = 1.43, $\gamma_{TA}$ = 1.26,$\gamma_{TA^{'}}$ = 1.22, $\gamma$= 1.60, $\theta_{LA}$ = 149.73 K, $\theta_{TA}$ = 148.64 K, $\theta_{TA^{'}}$ = 148.19 K, V = 19.05 $\times$ $10^{-30}$ $m^{3}$, and M = 196.69 $\times$ $10^{-27}$ kg.
The numerical values of various parameters used to calculate ($\kappa_L$) for Hf$_2$Ni$_2$InSb (tetragonal phase) are; $\nu_{LA}$ = 4963.8142, $\nu_{TA}$ = $\nu_{TA^{'}}$ =  2324.1295 m/s, $\gamma_{LA}$= 1.40, $\gamma_{TA}$ = 1.16,$\gamma_{TA^{'}}$ = 1.23, $\gamma$= 1.6, $\theta_{LA}$ = 148.13 K, $\theta_{TA}$ = 129.40 K, $\theta_{TA^{'}}$ = 123.89 K, V = 19.05 $\times$ $10^{-30}$ $m^{3}$, and M = 196.69 $\times$ $10^{-27}$ kg.

The numerical values of various parameters used to calculate ($\kappa_L$) for ZrNiSn are; $\nu_{LA}$ = 5322.6442 m/s, $\nu_{TA}$ = 3600.8698 m/s, $\nu_{TA^{'}}$ = 2852.1311 m/s, $\gamma_{LA}$ = 1.69, $\gamma_{TA}$ = 1.36,$\gamma_{TA^{'}}$ = 1.30, $\gamma$= 1.71, $\theta_{LA}$ = 213.76 K, $\theta_{TA}$ = 184.12 K, $\theta_{TA^{'}}$ = 166.74 K, V = 19.38 $\times$ $10^{-30}$ $m^{3}$, and M = 148.64 $\times$ $10^{-27}$ kg. The numerical values of various parameters used to calculate ($\kappa_L$) for Zr$_2$Ni$_2$InSb (cubic phase) are; $\nu_{LA}$ = 4784.2714 m/s, $\nu_{TA}$ = 4040.5217 m/s, $\nu_{TA^{'}}$ = 2774.3478 m/s, $\gamma_{LA}$ = 1.42, $\gamma_{TA}$ = 1.28,$\gamma_{TA^{'}}$ = 1.24, $\gamma$=1.60, $\theta_{LA}$ = 170.93 K, $\theta_{TA}$ = 171.02 K, $\theta_{TA^{'}}$ = 171.32 K, V = 19.37 $\times$ $10^{-30}$ $m^{3}$, and M = 148.40 $\times$ $10^{-27}$ kg.
The numerical values of various parameters used to calculate ($\kappa_L$) for Zr$_2$Ni$_2$InSb (tetragonal phase) are; $\nu_{LA}$ = 4696.1338 m/s, $\nu_{TA}$ = 3907.6455 m/s, $\nu_{TA^{'}}$ = 2588.8792 m/s, $\gamma_{LA}$ = 1.43, $\gamma_{TA}$ = 1.26,$\gamma_{TA^{'}}$ = 1.22, $\gamma$= 1.64, $\theta_{LA}$ = 166.19 K, $\theta_{TA}$ = 149.18 K, $\theta_{TA^{'}}$ = 142.16 K, V = 19.36 $\times$ $10^{-30}$ $m^{3}$, and M = 148.45 $\times$ $10^{-27}$ kg.

\subsection{Carrier Relaxation Time}

\begin{figure}[!htbp]
	\centering
	\includegraphics[scale=0.5]{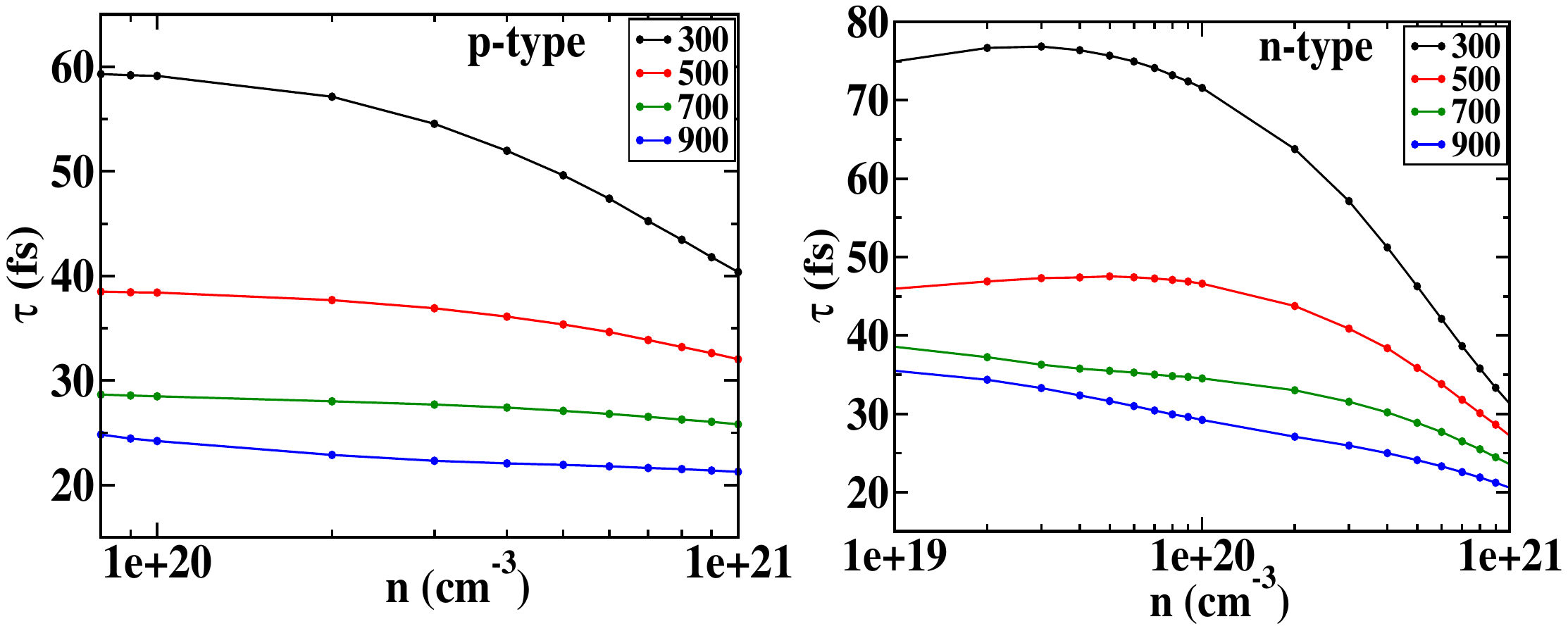}
	\caption{Temperature dependence of p-type (holes) and n-type(electrons) relaxation time for Cubic Hf$_2$Ni$_2$InSb.}
	\label{tau_Hf}
\end{figure}

\begin{figure}[!h]
	\centering
	\includegraphics[scale=0.5]{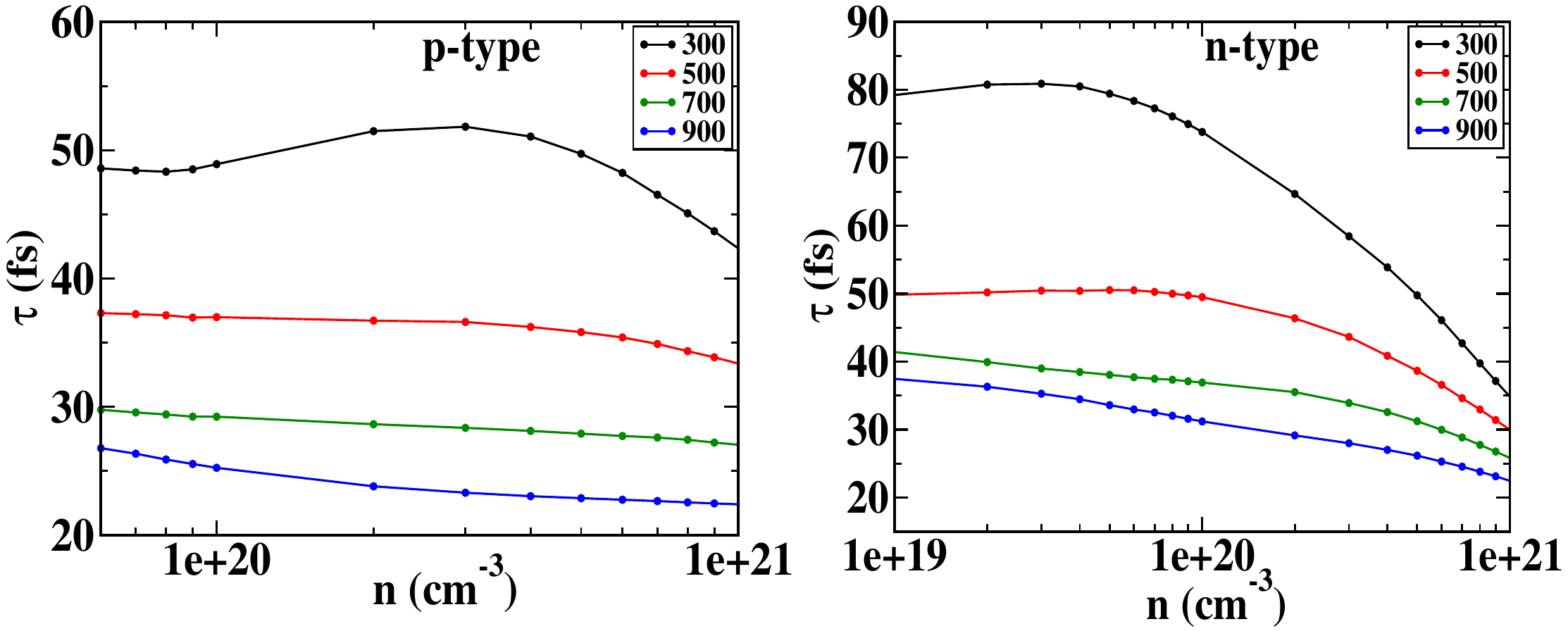}
	\caption{Temperature dependence of p-type (holes) and n-type(electrons) relaxation time for Cubic Zr$_2$Ni$_2$InSb.}
	\label{tau_Zr}
\end{figure}	

We have estimated the carrier relaxation time for electron and hole transport, using Ab-initio Scattering and Transport (AMSET)\cite{AMSET}. AMSET is a numerical code for calculating carrier relaxation time and transport properties within the ab-initio framework. Four scattering mechanisms are simulated in AMSET. They are  acoustic deformation potential scattering, ionized impurity scattering, polar optical phonon scattering and piezoelectric scattering. The mode dependent scattering rates are calculated using Born approximation. Based on Fermi's golden rule, the differential scattering rate from initial state $|$nk$\rangle$ to final state $|$mk+q$\rangle$ is calculated as:
\begin{equation}
\tau_{nk \rightarrow mk+q}^{-1} = \frac{2\pi}{\hbar}|g_{nm}(k,q)|^2\delta(\epsilon_{nk}-\epsilon_{mk+q}),
\end{equation}
where $g_{nm}(k,q)$ is the matrix element for scattering from state  $|nk\rangle$ into state $|mk+q\rangle$ and $\epsilon_{nk}$ is the energy of the state $|nk\rangle$.

The acoustic deformation potential (ADP) scattering is calculated by assuming that the lattice potential perturbation due to the thermal motion has linear dependence on relative volume change. The constant of proportionality is the deformation potential. A more general acoustic deformation potential (ADP) matrix element is given by:
\begin{equation}
g_{nm}^{ad}(k,q) = \sqrt{k_BT} \sum_{G\neq-q}^{}\bigg[\frac{\tilde{D}_{nk}:\tilde{S}_l}{cl\sqrt{\rho}}+\frac{\tilde{D}_{nk}:\tilde{S}_{t_1}}{ct_1\sqrt{\rho}}+\frac{\tilde{D}_{nk}:\tilde{S}_{t_2}}{ct_2\sqrt{\rho}}\bigg]\langle mk+q|e^{i(q+G).r}|nk \rangle
\end{equation}
where $\hat{S}$ = $\hat{q}\otimes\hat{u}$ is the unit strain for an acoustic mode, $\tilde{D}_{nk}$ = ${D}_{nk}$ + $v_{nk}$ $\otimes$ $v_{nk}$ in which $D_{nk}$ is the rank 2 deformation potential tensor, $\hat{u}$ is the unit vector of phonon polarization, and the subscripts $l$, $t_1$ and $t_2$ indicate properties belonging to the longitudinal and transverse modes.

In polar semiconductors, an electric polarization along with the deformation potential is produced by the lattice oscillations. Since the ions oscillate out-of-phase for optical modes, an electric dipole moment varying with time is generated. This leads to an additional interaction of optical phonons with charge carriers. The electric field due to perturbation of dipole moment between atoms leads to scattering of carriers. Thus, Polar optical phonon (POP) scattering rate is given by:
\begin{equation}
g_{nm}^{po}(k,q) = \bigg[\frac{\hbar\omega_{po}}{2}\bigg]^{1/2}\sum_{G\neq-q}^{}\bigg(\frac{1}{\hat{n}.\epsilon_{\infty}.\hat{n}}-\frac{1}{\hat{n}.\epsilon_{s}.\hat{n}}\bigg)^{1/2}\frac{\langle mk+q|e^{i(q+G).\boldsymbol{r}}|nk\rangle}{|q+G|}
\end{equation}
where $\epsilon_s$ and $\epsilon_{\infty}$ are the static and high-frequency dielectric tensors and $\omega_{po}$ is the polar optical phonon frequency.

The ionized impurity scattering is an important scattering mechanism at lower temperatures. The mobile carriers are attracted by ionized impurity which leads to screening of potential. The matrix element for ionized impurity scattering\cite{AMSET} is given by:
\begin{equation}
g_{nm}^{po}(k,q) = \sum_{G\neq-q}^{}{\frac{n^{1/2}_{ii}Ze}{\hat{n}.\epsilon_{s}.\hat{n}}}{\frac{\langle mk+q|e^{i(q+G).r}|nk \rangle}{|q+G|^2+\beta^2}}
\end{equation}
where $Z$ is defect charge, $n_{ii}$ is the concentration of ionized impurities which is equal to charge compensation $\times$ ($n_{holes}$-$n_{electrons}$)/$Z$ and $\beta$ is the inverse screening length.

The piezoelectric scattering (PIE) is basically polar acoustic phonon scattering. This scattering mechanism is important at very low temperatures. However we have included this scattering as well in our calculations.
In 300 to 900K, POP and ADP are the dominant scattering mechanisms. 





\section*{Acknowledgment}
BS acknowledges financial support from the Indian Institute of Technology, Bombay in the form of teaching assistantship. BS acknowledges Vikram for some initial discussions regarding the project. A.A. acknowledges DST-SERB (Grant No. CRG/2019/002050) for funding to support this research.

\section*{Author Contributions}  
BS and AA conceived the initial idea of the project. BS performed all the calculations and analyzed them with the help of AA. BS wrote the initial draft of the manuscript, which was further corrected by AA. AA supervised the entire project.

\section*{Competing financial interests}
The authors declare no competing financial interests.
\pagebreak
\end{document}